\documentclass[conference]{IEEEtran}
\IEEEoverridecommandlockouts
\usepackage{amsmath,amssymb,amsfonts}
\usepackage{amsthm}
\usepackage{algorithmic}
\usepackage{url}
\usepackage{textcomp}
\usepackage{xcolor}

\usepackage[caption=false,font=footnotesize]{subfig}
\usepackage[pdftex]{graphicx}

\usepackage{csquotes}

\DeclareMathOperator{\E}{\mathbb{E}}

\renewcommand{\P}{\mathbb{P}}

\newtheorem{theorem}{Theorem}
\newtheorem{corollary}{Corollary}

\newtheorem{lemma}{Lemma}
\newtheorem{problem}{Problem}

\def\bd{\begin{definition}}
\def\ed{\end{definition}}
\def\bt{\begin{theorem}}
\def\et{\end{theorem}}
\def\be{\begin{center}\begin{equation}}
\def\ee{\end{equation}\end{center}}
\def\bc{\begin{corollary}}
\def\ec{\end{corollary}}
\def\bl{\begin{lemma}}
\def\el{\end{lemma}}
\def\br{\begin{remark}}
\def\er{\end{remark}}
\usepackage{booktabs}
\usepackage{todonotes}

\begin{document}

\title{
Sharing is Caring: Analysis of Hybrid Network Sharing Strategies for Energy Efficient Multi-Operator Cellular Systems}

\author{
\IEEEauthorblockN{
Laura Finarelli,\IEEEauthorrefmark{1}\IEEEauthorrefmark{4}
Maoquan Ni,\IEEEauthorrefmark{2}
Michela Meo,\IEEEauthorrefmark{2}
Falko Dressler,\IEEEauthorrefmark{4}
and Gianluca Rizzo\IEEEauthorrefmark{1}\IEEEauthorrefmark{3}
}
\IEEEauthorblockA{\{laura.finarelli,gianluca.rizzo\}@hevs.ch,\{maoquan.ni,michela.meo\}@polito.it,dressler@ccs-labs.org}
%
\IEEEauthorblockA{
\IEEEauthorrefmark{1} HES SO Valais\;\;
\IEEEauthorrefmark{2} Politecnico di Torino\;\;
\IEEEauthorrefmark{3} Università di Torino\;\;
\IEEEauthorrefmark{4} TU Berlin
}
}

\maketitle

\begin{abstract}
This paper introduces a novel analytical framework for evaluating energy-efficient, QoS-aware network-sharing strategies in cellular networks. Leveraging stochastic geometry, our framework enables the systematic assessment of network performance across a range of sharing paradigms, including both conventional single-operator scenarios and advanced hybrid strategies that enable full integration and cooperation among multiple mobile network operators. Our framework incorporates diverse user densities, rate requirements, and energy consumption models to ensure comprehensive analysis. Applying our results to real-world datasets from French mobile network operators, we demonstrate that hybrid network sharing can yield substantial energy savings, up to $35\%$, while maintaining quality of service. Furthermore, our results allow us to characterizing how the benefits of network sharing vary as a function of the geographical and functional characteristics of the deployment area. These findings highlight the potential of collaborative sharing strategies to enhance operational efficiency and sustainability in next-generation cellular networks.

\end{abstract}

%
%
%

%
%
%

\section{Introduction}

The rollout of next-generation mobile networks (5G/6G) is propelled by the need to support increasingly stringent quality of service (QoS) requirements, including ultra-low latency, high data rates, massive device connectivity, and enhanced reliability \cite{xu2021survey}. Although 5G technology has seen widespread deployment, it continues to face significant challenges in meeting the growing demand for high-bandwidth and low-latency services in a resource-efficient manner. To address these requirements, Mobile Network Operators (MNOs) are increasingly deploying dense Radio Access Networks (RANs), which can lead to higher operational and capital expenditures (OPEX and CAPEX), increased energy consumption, and periods of infrastructure under-utilization, particularly in areas with fluctuating or uneven user demand \cite{oh2011toward}.

\begin{figure}
    \includegraphics[width=\columnwidth]{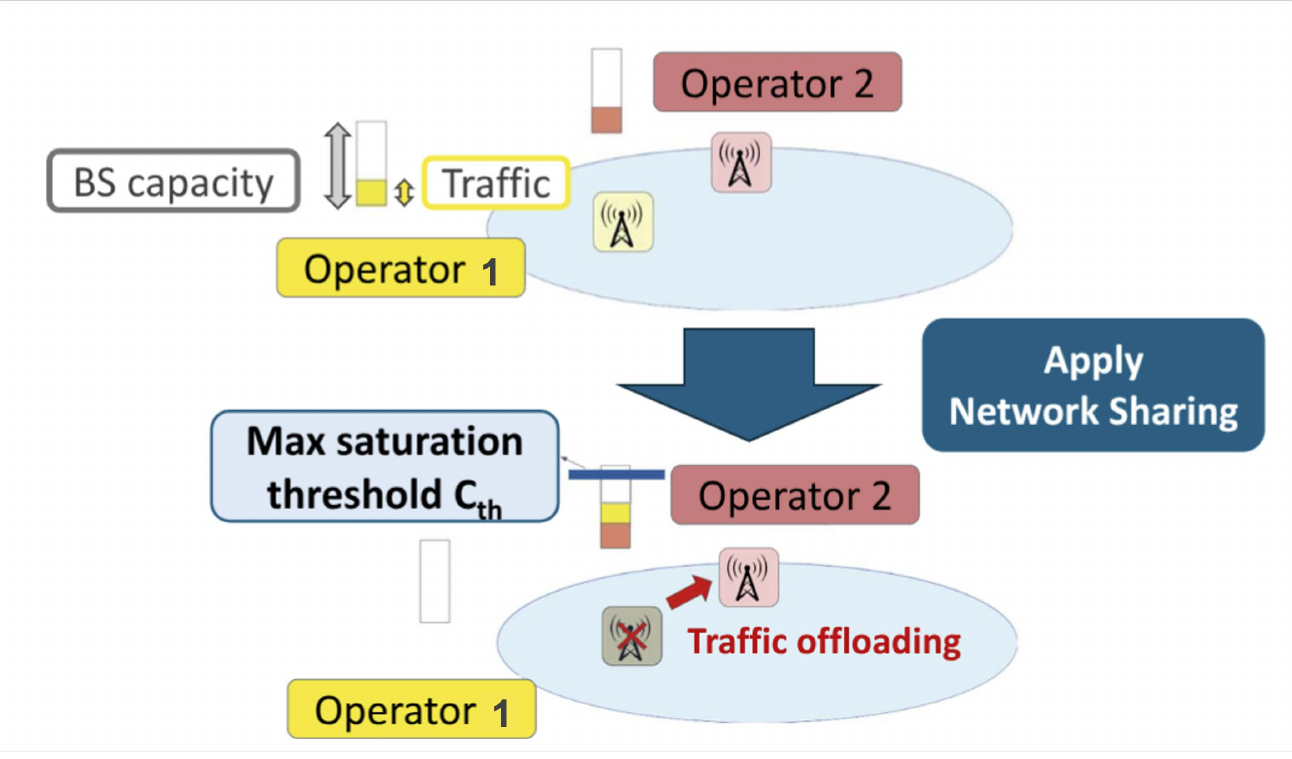}
    \caption{Scheme illustrating the idea underlying network sharing: when both the traffic of two operators is low, one can decide to transfer its load to the other. $C_{th}$ is a safety threshold beyond which offloading cannot take place.}
    \label{fig:ns_scenario}
\end{figure}

Under these circumstances, network sharing (NS) \cite{Xelshaer2018infrastructure,Xmahmoodi2017optimal} (Fig.~\ref{fig:ns_scenario}) has emerged as a potential strategy to tackle these inefficiencies. By allowing multiple MNOs to share base stations (BSs), NS enables more efficient use of network infrastructure. For instance, during off-peak hours, a BS can be placed in sleep mode while its traffic is seamlessly offloaded to a neighboring shared BS, reducing energy consumption and operational costs. Moreover, NS can significantly improve resource allocation and service continuity during emergencies or network failures, when only a subset of the infrastructure remains functional. This collaborative approach enhances both the resilience and efficiency of mobile networks, particularly under challenging or unpredictable conditions. \cite{RNDM} shows how two MNOs may achieve a fair, cooperative NS approach that lowers failure rates while promoting economic collaboration. 

Network sharing strategies in mobile communications range from basic infrastructure co-location to advanced resource virtualization \cite{Xelshaer2018infrastructure,Xmahmoodi2017optimal,Xmahmoodi2022infrastructure}. Active sharing enables multiple operators to share radio access equipment, either with separate spectrum (MORAN) or shared spectrum (MOCN). More advanced approaches include core network sharing, where operators share back-end systems, and spectrum sharing, which improves utilization of licensed frequencies \cite{Xmahmoodi2022infrastructure}. In 5G, network slicing introduces dynamic, virtual networks over shared infrastructure tailored to specific services. 
Previous research introduced data-driven NS strategies that dynamically adjust BS sleep modes, optimizing performance by tuning NS parameters according to traffic demands \cite{ICC}. 
Furthermore, dynamically configured network sharing strategies tailored to traffic and location-specific characteristics that can significantly optimize RAN efficiency are considered in \cite{SPAWC}. 

The impact of co-location on network resources was investigated in \cite{lópezpérez2024capacitypowerconsumptionmultilayer}, showing that it negatively impacts the network power consumption. 
\cite{leopectrumsharinglowearth} uses stochastic geometry to model spectrum sharing between terrestrial and non-terrestrial networks.
An analytical model for NS is proposed in \cite{andrews2017quantifyingbenefitsinfrastructuresharing} in which the strategies are investigated using game theory. However, the authors only focus on the cost of network sharing without any consideration of energy or QoS for served users. In \cite{farhat2017ran}, the authors investigated the optimal NS strategy, examining the profits for both users and MNOs without considering the energy constraints of the network.

Overall, existing studies on network sharing often overlook the spatial randomness of real-world deployments and rarely address both energy efficiency and QoS in a unified analytical framework. Most prior work focuses on cost or uses simulations, lacking generalizable, tractable models. Stochastic geometry enables rigorous, scalable analysis of diverse sharing strategies under realistic conditions. 

The present work fills a key gap, by providing an analytical tool for the joint evaluation of energy and QoS in networks where sharing strategies are applied. The advantage of our analytical approach lies in its ability to provide decisive information for network planning without being constrained to a specific setup.
In particular, our study presents a preliminary evaluation of energy savings achieved by synergistically combining traditional sleep mode mechanisms with a collaborative network-sharing strategy. This approach allows multiple operators to utilize their respective infrastructures collectively, resulting in improved overall network performance while maintaining fairness in resource distribution among all operators involved. In this perspective, the implementation of a cell-by-cell roaming strategy can be considered as well. 
To the best of our knowledge, this is the first work that applies stochastic geometry to model QoS-aware network sharing in a cellular network.
 
The main contributions can be summarized as follows:
\begin{itemize}
    \item We provide an analytical framework for a first-order evaluation of the potential energy savings of QoS-aware network sharing strategies.
    \item We apply it to a realistic, measurement-based scenario with two MNOs, and we evaluate the savings achievable by traditional NS approaches, based on switching off the whole network of one or more operators, as well as those attainable through a collaborative approach by which all operators share resources and infrastructures.
    Results show that energy savings achievable in such realistic scenarios reach up to $35\%$ with respect to QoS-aware sleep modes applied independently by each operator.
\end{itemize}

\section{System Model}

We consider a region of space in which a set of base stations (BSs) are deployed, modeling a 5G/6G cellular network. Each BS belongs to one of $I$ MNOs. 
A fraction  $c$ of the overall base stations are co-located, i.e., they share the installation site (and thus the same position in space) with BSs from other operators. This models deployments in which several MNOs share the same antenna mast or roof area. In what follows, we assume that at every point in space in which there is colocation, there is a BS from each of the $I$ MNOs, though our analysis can be extended easily to more general settings.  
BSs are assumed to be Urban Macrocells (UMs) \cite{haneda20165g}. Note, however, that our framework can be easily extended to any combination of UMs, small cells, and femtocells.
For any MNO $i$, users are distributed in space according to a homogeneous Poisson Point process (PPP) with intensity  $\lambda_u^i$. 

For each MNO $i$, we model the BS locations as a homogeneous PPP with spatial intensity $\lambda_i$. 
For ease of analytical treatment, we adopt a channel model that considers only distance-dependent path loss. However, our findings can be easily extended to include other propagation effects, such as stochastic fading and shadowing. We assume random frequency reuse is in place.
We focus on the downlink component of the cellular network, as it accounts for the majority of energy consumption at the BS level \cite{lópezpérez2024capacitypowerconsumptionmultilayer}. We assume that every user is served by the BS that provides the largest signal-to-noise ratio (SINR) at the user's location. Due to interference, such BS may not necessarily be the nearest one. Nevertheless, in settings with a significant propagation attenuation (i.e., when the attenuation coefficient \(\alpha\) is $3$ or higher, which is often the case in urban areas), this assumption remains a reasonable approximation \cite{bacelli}.
According to Shannon's capacity law, the capacity of a user of the $i-th$ MNO located at a distance \(r\) from its serving BS is 
\[
C_i(r) = \frac{B_i}{k} \log_{2}\left(1 + \frac{P r^{-\alpha}}{N_0 + I_i(r,k)}\right)
\]
Here, \(B_i\) stands for the channel bandwidth,  \(I_i\) is the total interference power received from the other BSs of the $i-th$ MNO, \(k\) is the reuse factor, $P$ the BS transmit power, and \(N_0\) is the power spectral density of the additive white Gaussian noise. As each MNO uses a different frequency band, the interference perceived by a user is due only to BSs from the same MNO as the serving BS. \\
To model the QoS perceived by a user, we use the \textit{per-bit delay}, defined as the inverse of short-term user throughput \cite{delay1}. 
Specifically, the key performance parameter of the network is the Palm expectation of the per-bit delay experienced by a typical user who is just beginning service \cite{rengarajan2015energy}.\\
We assume users are partitioned into $J$ classes, and let $j$ be the label of the $j-th$ class. This partition is made according to the required QoS, in terms of mean per-bit delay $\tau^0_j$ (or equivalently, of mean short-term throughput $R_{0,j}=(\tau^0_j)^{-1}$).
$\forall j$, $\gamma_j$ denotes the mean fraction of users belonging to class $j$. 
We assume that a weighted processor-sharing (WPS) mechanism is employed to allocate BS time among all connected users. For any user class $j$, let $w_j$ be the WPS weight associated with that class, with $w_J=1$.
If $N_j$ is the number of users from class $j$ at a given BS, then the utilization of that BS (i.e., the fraction of time during which it is actively serving users) is   
$U=\frac{\sum_j w_jN_j}{\sum_j w_jN_j+\iota}$, where $\iota$ is the fraction of time during which the BS does not serve any user. Thus, by tuning $\iota$, each BS may vary the amount of time spent serving users, and thus the QoS perceived by each of them. The fact that all users from the same class have the same WPS weight implements a degree of fairness among users within the same class and associated with the same BS, as they all receive an equal share of BS time. At the same time, to ensure fairness in resource scheduling between classes, we assume that $\forall j$, $w_j=\frac{R_{0,j}}{R_{0,J}}$. 
In what follows, we assume that each BS in the network tunes its utilization in such a way as to have, for each user class, the Palm expectation of the per-bit delay perceived by all the users coincide with the target value for that class.
As for the BS energy consumption, we consider a flexible, measurement-based model, by which the power consumed by a BS with utilization $U$ from MNO $i$ is \cite{boi@swipt}:
\begin{equation}\label{eq:energymodel}
E_i(U,P) = q_1 + U (q_2+q_3 P)
\end{equation}
$q_1$ models the part of the consumed power that is independent of the traffic load and transmit power. $q_2$ and $q_3$ model the contribution to power consumption due to computational and cooling processes, which scale proportionally with utilization, as well as the power consumed during signal transmission, which depends on both utilization and transmit power.

To save energy, we assume that each operator $i$ may turn to low power mode (or completely off) a fraction $(1-\beta_i)$ of its BSs. We assume random sleep modes are used, by which each BS from the $i-th$ operator may be turned off with probability $(1-\beta_i)$. 

\section{User-Perceived Performance}

In this section, we derive analytical expressions for the primary performance indicators of our system as a function of its main parameters, based on the tools of Stochastic Geometry. To this end, we focus on the derivation of the \textit{ideal} per-bit delay experienced by the typical user, defined as the per-bit delay experienced when the serving base station has utilization equal to 1. Thus, the \textit{actual} per-bit delay perceived by the typical user is given by the product of the ideal per-bit delay and the utilization of its serving BS.

Let $S(x)$ denote the location of the BS that is serving the user located at $x$. For the typical class $j$ user at the origin, served by the BS  in $S(0)$ from the $i-th$ operator, the ideal per-bit delay perceived is
\begin{equation}\label{eq:tau_sbs}
   \tau_j^i(S(0)) = \frac{\sum_j w_jN_{j}}{ w_jC_i(D(0))} 
\end{equation}
where 
 $D(0)$ is the distance between the given user at the origin and its serving BS.

For the derivation of an analytical expression for the Palm expectation of the ideal per-bit delay, we need to derive the average interference perceived by a user. Let $\bar{\tau}_j^i$ denote the Palm expectation of the ideal per-bit delay perceived by a typical class $j$ user joining a network where network sharing is in place, and served by a BS from the $i-th$ MNO. We have the following result:
\begin{lemma}[Mean Interference power]\label{lemma:interf_hetnet}
The mean interference perceived by a user of the $i-th$ MNO at a distance $r$ from its serving BS is:
\begin{equation*}
    \bar{I}_i (r) = 
    \frac{2 P \pi r^{2-\alpha} }{\tau^0_J k(\alpha-2)} \bar{\tau}_J^i \beta_i\lambda_i 
\end{equation*}
\end{lemma}
\textbf{The proof is an extension of the one for Lemma 1 in \cite{rizzo2018energy}.}
The following theorem expresses the Palm expectation of the ideal per-bit delay $\bar{\tau}_j$ as a function of the main system parameters. 

\begin{theorem}\label{th:mean_tau}
    In a system with $I$ MNOs, when network sharing is in place, for any user class $j$ the Palm expectation of the ideal per-bit delay perceived by a typical class $j$ user joining the system, 
    is the unique solution of the following fixed-point problem:
    \vspace{-0.05in}  
\begin{equation}\label{eq:tau_moving}
        \bar{\tau}_j  = \sum_i\bar{\tau}_{j}^i p_i
  =  \sum_i p_i\lambda_{u}^i(\sum_j \gamma_jw_j)  \int_0^\infty h(r)H_i(P,r) dr
\end{equation}
where $p_i = \frac{\lambda_c}{\sum_{i'}\beta_{i'}\lambda_{i'}} p_c + \frac{\beta_i\lambda_i-\lambda_c/I}{\sum_{i'}\beta_{i'}\lambda_{i'}-\lambda_c}(1 - p_c)$ is the probability of being served by MNO $i$, $p_c$ is the probability of a BS to be co-located, $\beta_i \lambda_i$ is the fraction of active BS of operator $i$, and
\[
H_i(P, r) = 
\begin{aligned}
    &\frac{e^{-(\sum_{i'} \beta_{i'}\lambda_{i'})\pi r^2}
    (\sum_{i'} \beta_{i'}\lambda_{i'}) 2\pi r}
    {C_i(r)}
\end{aligned}
\]
 \begin{equation}\label{eq:h_r}
h(r)=
\int_0^\infty \int_0^{2\pi} e^{-\sum_{i'}\beta_{i'}\lambda_{i'} A(r,x,\theta)}x d\theta dx
\end{equation}
and $A(r, x, \theta)= \pi x^2 - 
\bigg[r^2 \arccos\left(\frac{r +  x \sin(\theta)}{d(r, x, \theta)}\right)+$

$+x^2\arccos \left(\frac{x + r \sin(\theta)}{d(r, x, \theta)}\right)-\frac{1}{2} \sqrt{r^2-(d(r, x, \theta)-x)^2 }\cdot$
$\cdot\sqrt{(d(r, x, \theta) +x)^2-r^2} \bigg]$. 
$d(r,x,\theta)$ being the Euclidean distance between $(x,\theta)$ and $(0,-r)$.
\end{theorem}
\begin{proof} 
 The Palm expectation of $\tau_j^i(S(0),D)$ perceived by a user in $S(0)$ (i.e. at the origin) at a distance $D$ from its serving BS is $\E^0[\tau_j(S(0),D)] = \sum_i\E^0[\tau_j(S(0),D) \vert i] p_i$,
where the conditioning is to the case in which the typical user is served by a BS from MNO $i$, and $p_i$ is the corresponding probability. 
Then $\E^0[\tau_j(S(0),D) \vert i]=\E^0[\tau_j^i(S(0),D)]= \bar{\tau}_j^i$. 
Thus, $\E^0[\tau_j^i(S(0),D)]=\E^0\biggl[\frac{\sum_j w_jN_j(S(0))}{C(D(0), P,I,B_i)} \biggr] \approx \int_0^\infty \frac{ 1}{ C_i(r,P,\bar{I}(D),B_i)}\cdot$
\[\cdot\E^0\biggl[\sum_j w_jN_j(S(0))  \vert r \leq D \leq r+dr\biggr] P(r\leq D\leq r+dr)dr\]
where $\bar{I}(D)$ is the mean interference given by Lemma \ref{lemma:interf_hetnet}. For $dr \rightarrow 0$ we have $P(r \leq D \leq r + d r) \approx pdf_R(r)$, where $pdf_R(r) = e^{-(\sum_{i'} \beta_{i'} \lambda_{i'})\pi r^2}
(\sum_{i'} \beta_{i'} \lambda_{i'}) 2\pi r$ is the probability distribution function of the distance of each user from its serving BS.

To derive the final expression, we compute the expected value of the Poisson distribution of users served by a BS from $i$ $\E^0[\sum_j w_jN_j(S(0)) |r \leq D \leq r + d r ]$  of intensity $\lambda_{u}^i (\sum_j \gamma_jw_j)$. To compute an expression for the average size of the Voronoi cell $h(r)$,  we move the typical user in $(0,-r)$ so that its serving BS is located at the origin. Then we consider a user in $(x,\theta)$ served by the BS at the origin and impose that no other BSs are closer. This event, using the void probability of the superposed process of all BS with intensity $\sum_{i'} \beta_{i'} \lambda_{i'}$, occurs with a probability $e^{-(\sum_{i'}\beta_{i'} \lambda_{i'})A(r,x,\theta)}$, where $A(r,x,\theta)$ is the area of the circle centered at $(x,\theta)$ that is not overlapped by the circle centered at the origin. To derive the probability $p_i$, we should account for the existence of co-located BSs. 
To this aim, we define the thinned PPP of co-located BS, which has intensity $\lambda_c = c (\sum_{i'} \beta_{i'}\lambda_{i'})/I$ and the thinned PPP of BS from $i$ that are not co-located which has intensity $\Tilde{\lambda}_i = \beta_i\lambda_i - \lambda_c/I$. It holds that $\sum_{i'} \beta_{i'}\lambda_{i'} = \sum_{i'}\Tilde{\lambda}_{i'} + \lambda_c $
Specifically: 
$$p_i = \P(R_i \leq R_{-i} \vert C) p_c + \P(R_i \leq R_{i'} \vert \bar{C}) (1-p_c)$$
where $p_c$ is the probability of being co-located, given by the fraction $c$ of co-located BS, $C$ is the event: \textit{base stations are co-located}, $R_i$ and $R_{-i}$ are the RV for the distances between users and tier $i$ and all the other tiers expect from $i$, respectively. We proceed with the derivation of $\P(R_i \leq R_{-i} \vert C)$ because the remaining probability can be obtained following the same steps and using the thinned process of non-co-located BS.
$$\P(R_i \leq R_{-i} \vert C) = \int_0^\infty \P(R_{-i} > r) pdf_{c}(r)dr$$
where $pdf_c$ is obtained in a canonical way as the pdf of the distance between the typical user and the thinned PPP of co-located BS, which has intensity $\lambda_c$. The integral becomes:
$$\int_0^\infty e^{-(\sum_{i'} \lambda_{i'}-\lambda_c)\pi r^2} 2 \pi r \lambda_c  e^{-\pi r^2 \lambda_c}dr = \frac{\lambda_c}{\sum_{i'} \lambda_{i'}}$$
Finally:
$$p_i = \frac{\lambda_c}{\sum_{i'}\beta_{i'}\lambda_{i'}} p_c + \frac{\beta_i\lambda_i-\lambda_c/I}{\sum_{i'}\beta_{i'}\lambda_{i'}-\lambda_c}(1 - p_c)$$\\
The existence and uniqueness of the fixed point equations for $\bar{\tau}_j^i$, derive from applying the Banach fixed-point theorem to the problem at hand, as by using well-known inequalities, it can be proved that the system of equations for the average per-bit delays is a contraction.\end{proof}
As a corollary, note that, when there is no sharing, the ideal per bit delay for a class $j$ user from the $i-th$ MNO is given by the expression for $\bar{\tau}_{j}^i$ in Theorem 1.

\section{Energy-optimal Full Network Sharing}
Theorem 1 gives an analytical expression for the ideal per-bit delay perceived by a user, as a function of the density of active BS for each operator. When no sharing is in place, the tuning of the fraction of active BSs is performed independently by each operator, with the goal of delivering the target QoS to its users while minimizing the energy consumed. 
In this section we consider a version of NS, that we denote as \textit{full NS}, 
in which every MNO may serve users from other MNOs, and every MNO agrees to tune the fraction of its BS which are active in order to minimize the total amount of energy required to serve the aggregate of all users, while delivering the target QoS to all the users from all classes and MNOs. Thus, in the Full NS scheme, the fraction of active BSs for each MNO determines the service capacity which each MNO provides to the aggregate of all users from all MNOs participating to the NS scheme. 

In this section, we formulate the energy-optimal problem, which aims to identify for each MNO the optimal fraction of active infrastructure which allows delivering the target QoS to all the users from all classes and MNOs, while minimizing the energy required to perform this task. The input for this problem includes, for each operator $i$, the density of deployed BSs $\lambda_i$, and the density of users from each MNO.

\begin{problem}{\textbf{Energy optimal network configuration for full NS}}\label{Prob:1}
\[
\underset{\{\beta_i\}}{\text{minimize}}\   \sum_i \beta_i\lambda_{i}E(\bar{\tau}_i,P)
 \]
\vspace{-0.1in} Subject to:
\begin{align}
   \forall t,j\quad\bar{\tau}_j \leq \tau_j^0 
   \label{constraint}\\
   \forall i,\quad 0\leq \beta_i\leq 1
\end{align}
\end{problem}
where $E_i$ is the energy consumed by a BS of the $i-th$ MNO, from Eq. \ref{eq:energymodel}, and $\bar{\tau}_j$ derives from Theorem 1. We assume that operators are \textit{self sufficient}, i.e., that in absence of network sharing, for every MNO $i$, there exist at least one value of $\beta_i$ for which the operator is able to provide its users with the required QoS. It is easy to see that, when every operator is self sufficient, Problem 1 has always a solution.  

The problem is linear in the variables but has non-linear constraints. However, it has been solved optimally using the interior point method with nonlinear constraints. 

%
\section{Numerical Evaluation}
\label{evaluation}
In this section, we assess numerically our analytical approach to energy-optimal NS and characterize the optimal sharing strategies stemming from the solutions of Problem 1.

\begin{figure}
\begin{centering}
\includegraphics[width=0.8\columnwidth]{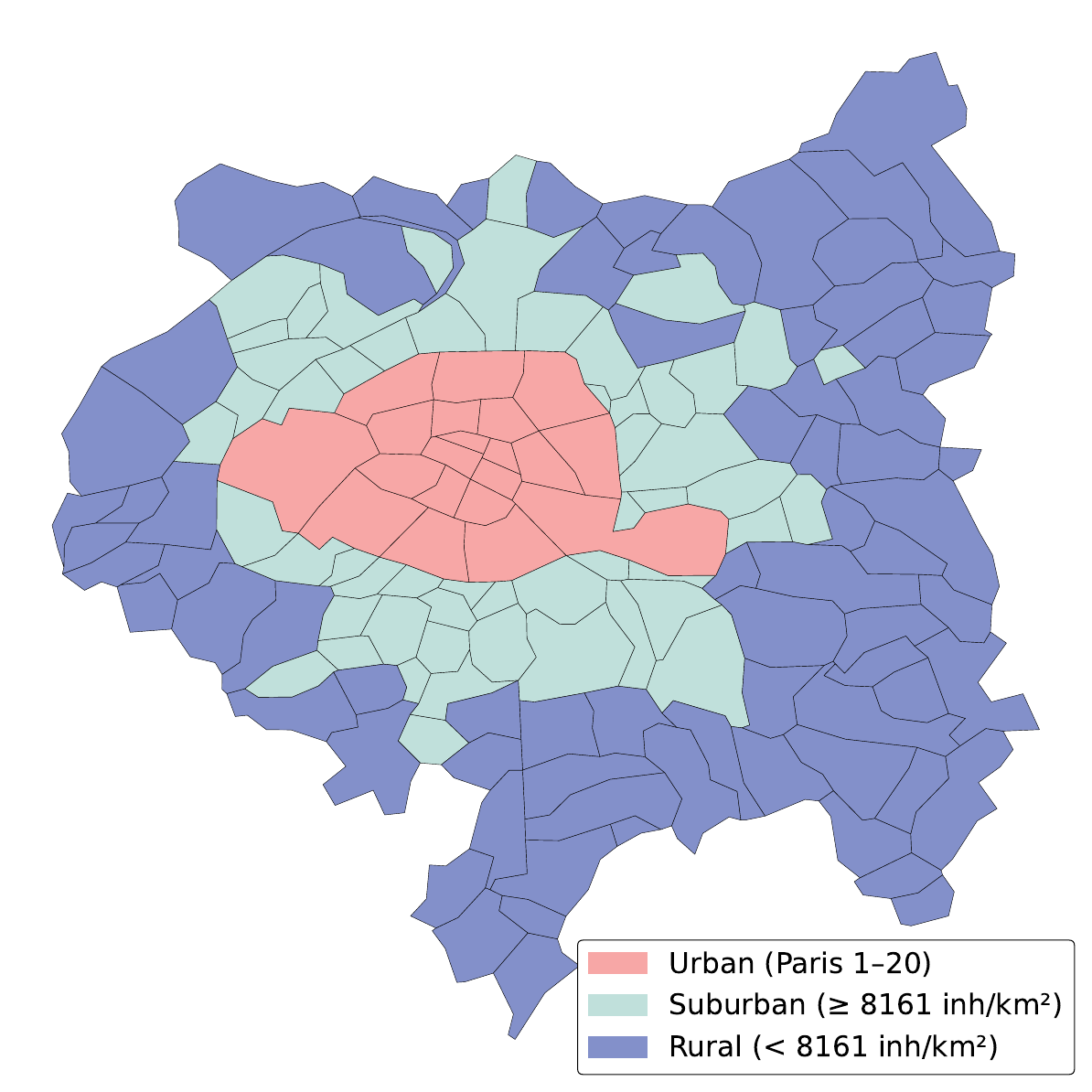}
    \caption{The considered region of Paris, partitioned into urban, suburban, and rural areas.}
    \label{fig:area}
\end{centering}
\end{figure}

\subsection{Setup}
We considered two different setups for the BS energy model. The first, denoted as HLP (high load proportionality), models more recent macro BS architectures characterized by high modularity, which allows a power consumption that is more sensitive to the amount of traffic served by the BS. Specifically, we considered a fixed energy consumption accounting for $36\%$ of the overall energy consumed \cite{boi@swipt}. The second class of BSs, denoted as LLP (low load proportionality), is characterized by a fixed component accounting for the vast majority of the overall energy consumption (set to $75\%$ \cite{boi@swipt}). In both cases, we assumed a transmit power of $20$ W, and a bandwidth of $20$ MHz \cite{band}.
In our evaluation, in addition to full NS, we consider the following network sharing strategies:
\begin{itemize}
	\item \textit{No sharing}: In this scenario, each MNO serves only its subscribers independently of the other MNOs, thus without sharing its network resources. Each MNO adopts sleep strategies to adapt the fraction of its active BSs to the actual demand.
	\item \textit{Operator switch off}: Only the network of one operator is active, and it serves users from all operators. 
\end{itemize}


\begin{figure*}
    \centering
     \subfloat[rural]{\includegraphics[width=.32\linewidth, height=2in]{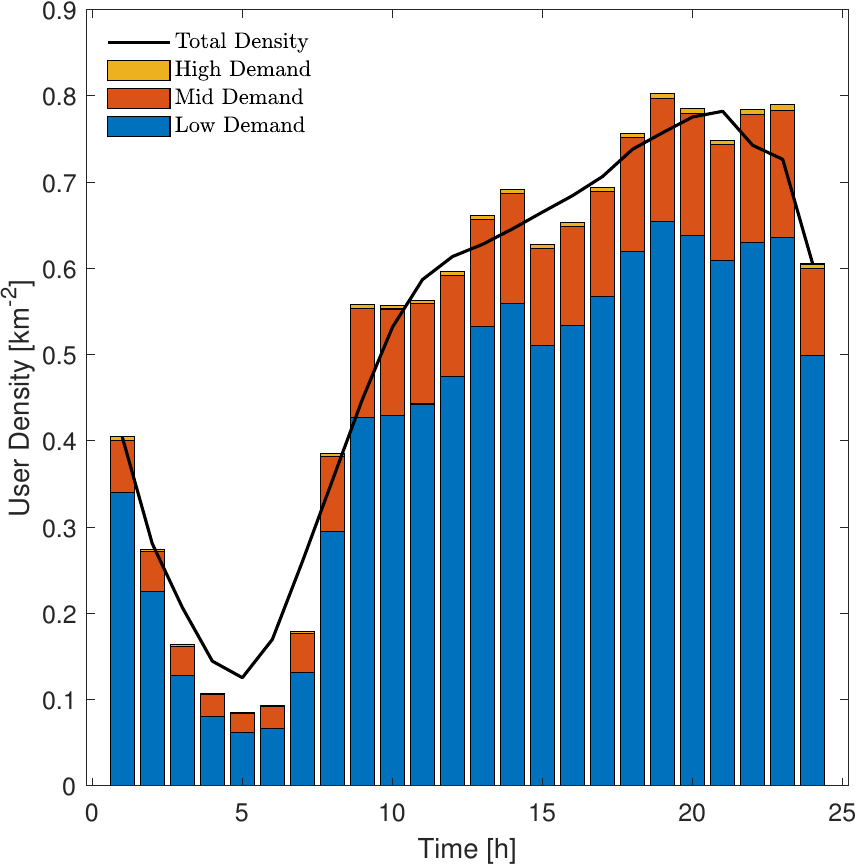}}
     \subfloat[urban]{ \includegraphics[width=.32\linewidth, height=2in]{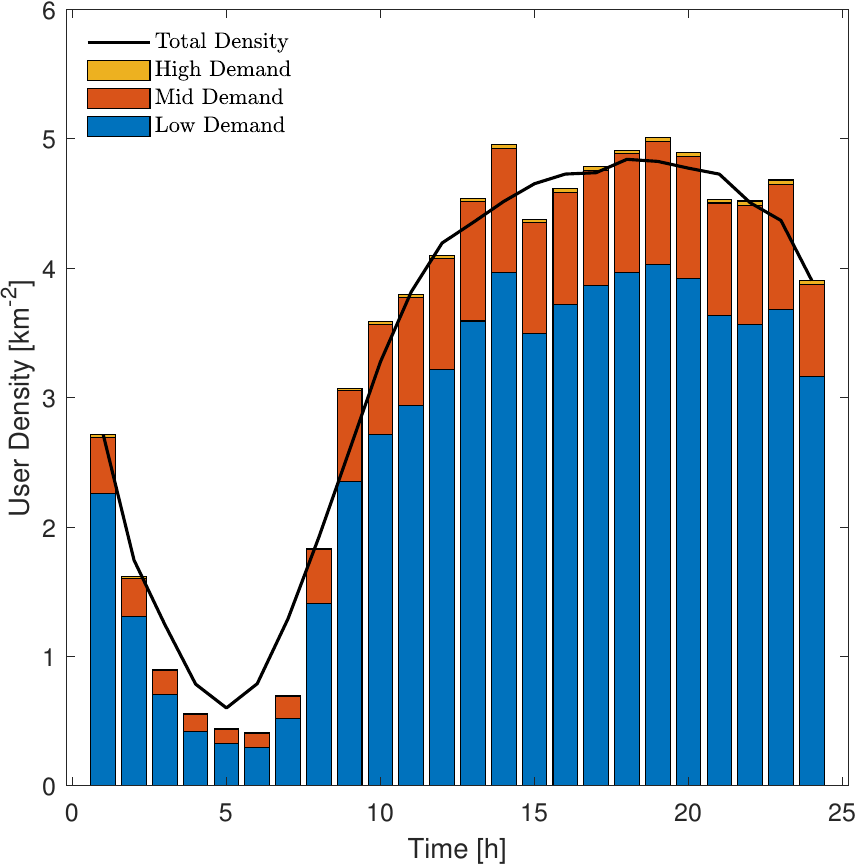}}
     \subfloat[suburban]{\includegraphics[width=.32\linewidth, height=2in]{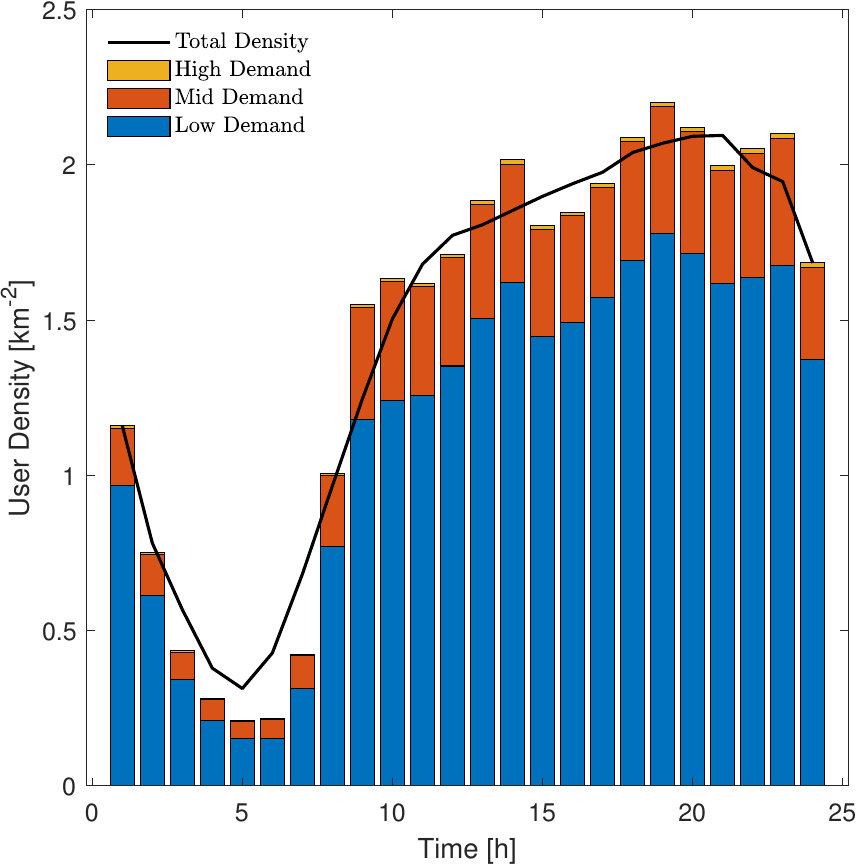}}
     \caption{Mean traffic profile over $24$ h per traffic class, for the Paris scenario, for the three area types.} 
     \label{fig:profiles}
\end{figure*}
\begin{figure*}
    \centering
     \subfloat[weekday]{\includegraphics[width=.35\linewidth, height=2in]{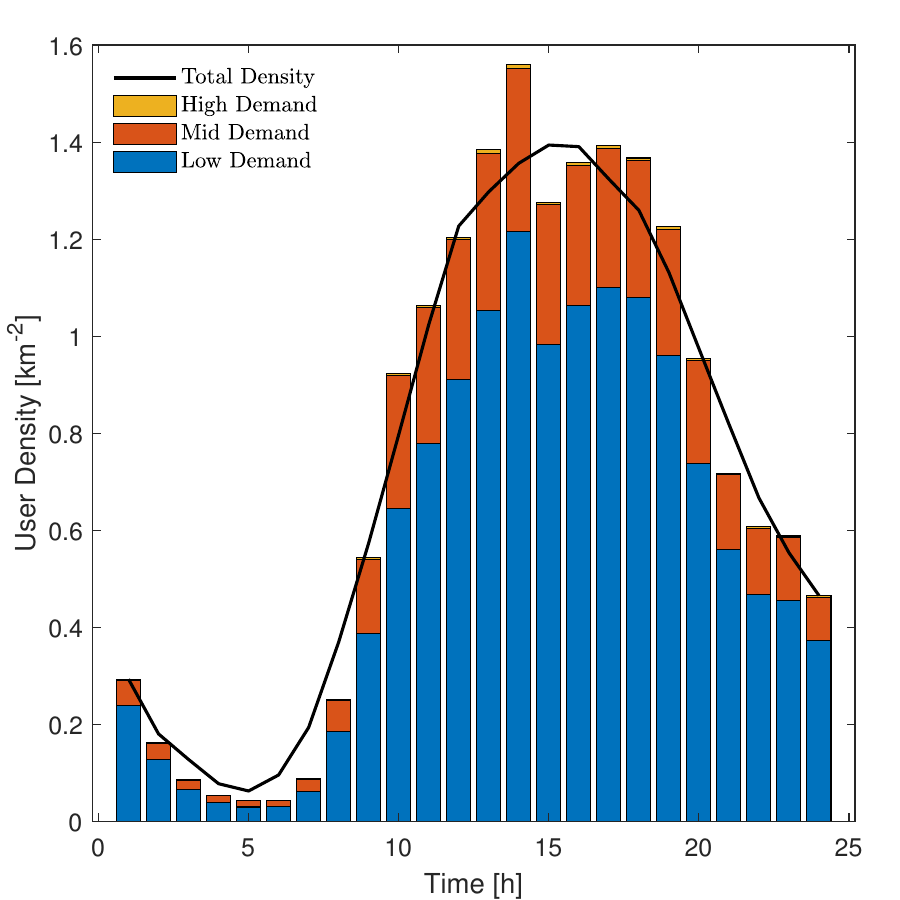}}
     \subfloat[weekend]{ \includegraphics[width=.35\linewidth, height=2in]{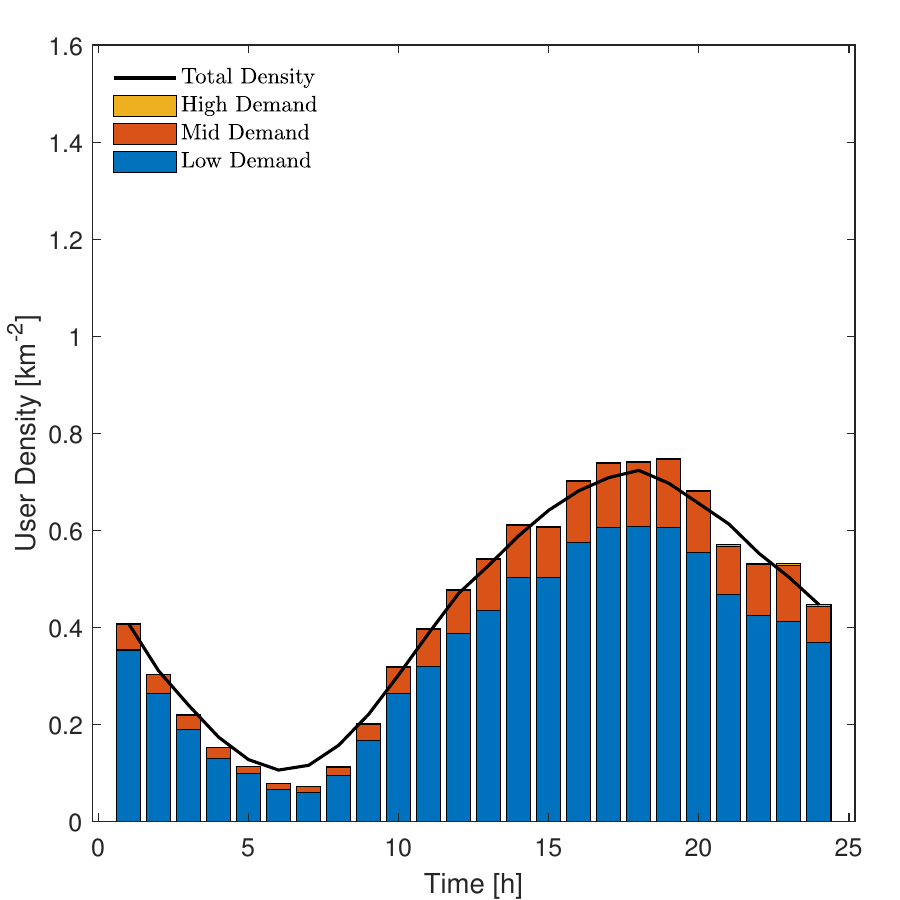}}
     \caption{Mean business traffic profile over $24$ h per traffic class, for the Paris scenario, per (a) weekday and (b) weekend.
     } 
     \label{fig:profiles_business}
\end{figure*}

To evaluate the effectiveness of our analytical approach, we applied it to a measurement-based scenario constituted by the urban region of Paris. Specifically, data are extracted from \cite{netmob23} and are relative to two of the main French MNOs.
The position of each BS for the two MNOs, as well as the fraction of colocated BS for each operator, are derived from \cite{ANFR}.

The density of users for each network has been derived from traffic volume, as follows.
We have categorized the traffic volume into three classes: \textit{high-demand}(H), which includes video streaming services, with a target mean rate of $5$ Mbit/s; \textit{medium-demand} (M), relative to audio streaming applications, with a mean rate of $0.2$ Mbit/s; and \textit{low-demand} (L), comprising social media, business apps, and general web browsing, with a $0.05$ Mbit/s rate. 
Generally, in realistic scenarios, user densities for each user class vary over time. However, many of these variations are known to exhibit periodic behavior over different timeframes (e.g., days, weeks). Thus, in our analysis, we consider the given urban region over a $24$ h period \cite{netmob23}. Such a time interval has been divided into slots of $15$ min, a duration which is generally short compared to the speed at which user density and network traffic vary over time. Thus, within each slot, we have assumed user density to be constant in the scenario.

For the urban area under study, we have derived the share of each traffic class over each 15-minute interval in the given period. To derive an estimation for the mean number of users for each class at each BS and at each time slot, for each class, we have divided the traffic volume by the mean data rate of that class. Then, through Voronoi tessellation, we have derived the spatial distribution of users in the whole region, for the three classes and for each time slot in the $24$ h period.  

To get insights into such a spatial distribution, we have partitioned the districts of the Paris region into urban/suburban/rural (Fig. \ref{fig:area}), as follows. The twenty arrondissements of Paris are classified as urban. For the remaining districts, we adopted INSEE’s median population density ($8161$ inhabitants per km² \cite{INSEE}) as a threshold between suburban and rural regions. 

Figure \ref{fig:profiles} displays daily traffic patterns for urban, suburban, and rural areas for the three classes of traffic. All of the regions follow the same daily pattern, which is linked to people's daily actions: morning increases, lunch peaks, and evening peaks associated with relaxing time. The only significant variation is scale: urban areas have the highest traffic since they have the most users; suburban areas are in the middle, and rural areas show the lowest volumes as well as the user density. The fraction of BS which are colocated in the scenario is roughly the same in all of the three type of areas in the Paris scenario ($31.980$ in urban areas, $27.810$ in suburban areas, and $27.850$ in rural areas). 

To identify traffic profiles linked to a business/office area use, if \(V_{\max,\mathrm{we}}\) and \(V_{\max,\mathrm{wd}}\) denote the weekend and weekday traffic peak for a BS, respectively, we define its traffic profile as business if
$\frac{V_{\max,\mathrm{we}}}{V_{\max,\mathrm{wd}}} < P_{\mathrm{th}}$
with \(P_{\mathrm{th}}=0.6\). This criterion captures BSs that are relatively idle on weekends, thus offering a high potential for energy savings via NS schemes during weekends. As a result, $6.4\%$ of all BSs in the Paris region have been classified as having a traffic profile of the business type. Figure \ref{fig:profiles_business} shows the daily traffic patterns of the business type, for the three classes of traffic. Business traffic profile exhibits a narrower peak of traffic with respect to the general traffic profile of the urban area, and a peak-to-trough ratio that goes from $14$ during weekdays to about $7$ over weekends.  
%
%
%

\subsection{Assessment of Network Sharing strategies }
In a first set of experiments, we have derived for every time slot, the energy consumed by the full NS strategy (Problem 1), as well as that consumed by the operator switchoff strategies, and we have compared the overall energy savings of each of these strategies with respect to the case of no sharing, in which each operator dynamically adapts its fraction of active BSs to variations in user density and relative share of user classes, for its own users.

\begin{figure*}
    \centering
     \subfloat[HLP energy model]{\includegraphics[width=.4\linewidth]{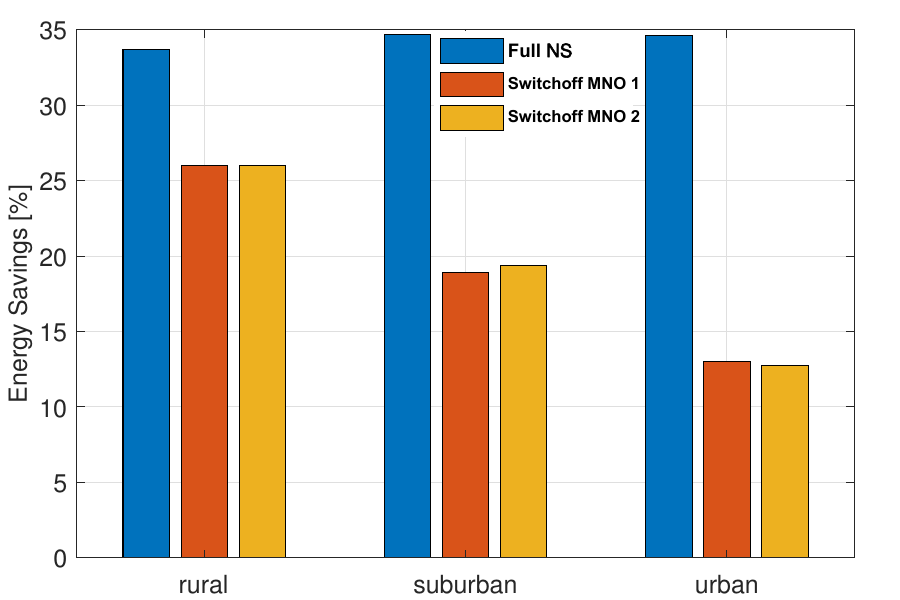}}
     \subfloat[LLP energy model]{\includegraphics[width=.4\linewidth]{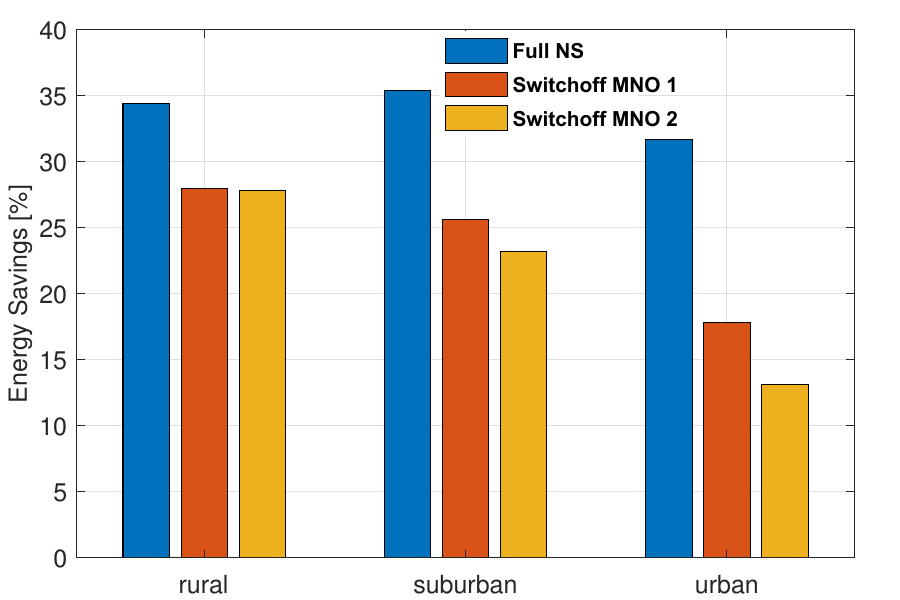}}
     \caption{Energy savings over the $24$ h of the considered network sharing strategies with respect to the energy consumed when each MNO independently applies sleep modes to optimize its energy consumption without network sharing, for the Paris scenario.} 
     \label{fig:energy_savings_noclasses}
\end{figure*}

Figure \ref{fig:energy_savings_noclasses}
 shows that in all areas, all sharing strategies allow achieving substantial energy savings with respect to the configuration in which every MNO optimizes its energy consumption independently.
In all settings, the largest savings are achieved by the full NS strategy, in spite of a higher implementation complexity. The savings are larger in areas characterized by a higher peak traffic, and thus by a larger amount of overdimensioning of network resources. In urban areas, in particular, the full NS strategy almost doubles the savings with respect to more traditional approaches based on operator switchoff. 
As visible in the figure, the combination of NS strategies with BS architectures with high load proportionality has a synergetic effect on energy savings, as it increases the potential impact that the dynamic tuning of active BSs may have on the overall energy consumption of the network.

\begin{figure*}    
 \centering
	\subfloat[]{\includegraphics[width=.40\linewidth]{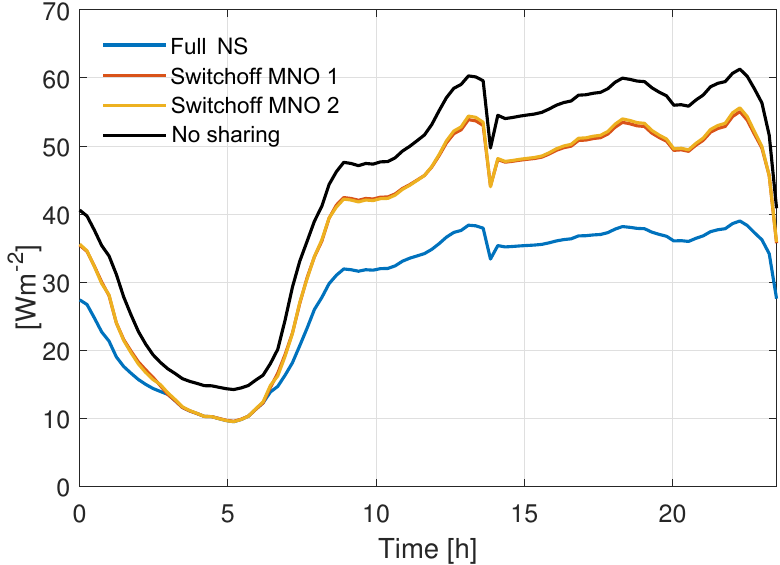}} 
    \hspace{0.01\linewidth}
     \subfloat[]{\includegraphics[width=.40\linewidth]{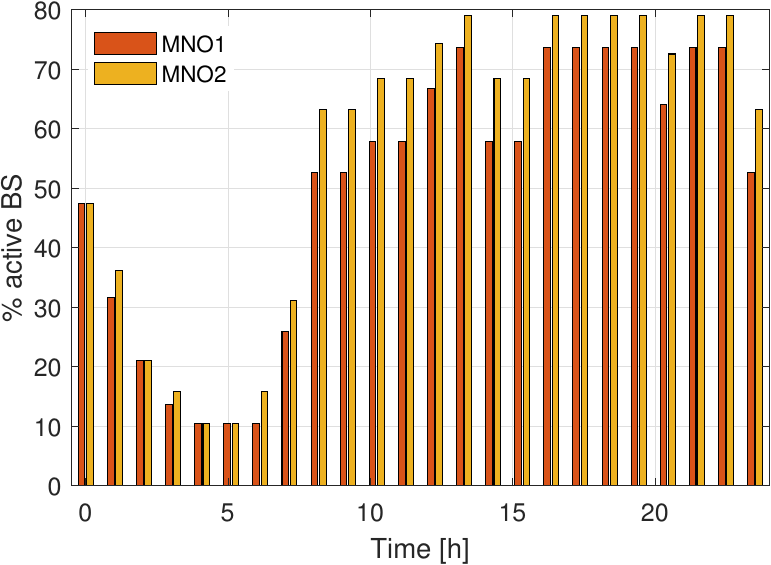}}
	\caption{(a) Energy consumed by the different NS strategies, as a function of the time slot over the 24 h, for the HLP BS energy model (b) Percentage of active base stations at the optimum for each MNO as a function of the time slot over the 24 h, for the full NS strategy.} 
	\label{fig:urban_daily}%
\end{figure*}

Figure \ref{fig:urban_daily}(a) shows the amount of energy consumed in each time slot for the different strategies, for the given scenario. 
Despite the fact that network sharing schemes are often considered as a way to save energy in periods of low loads, the plot shows that for all strategies, the most significant savings are achieved when traffic peaks, with the full NS scheme achieving more than $30\%$ reduction over energy-optimal network management without sharing.

Figure \ref{fig:urban_daily}(b) shows the percentages of active infrastructures per operator and time slot at the optimum, for the full NS scheme. 
This plot shows that the service load, in terms of fraction of active BSs, is almost equally distributed between the two operators.
This is expected, as in the considered scenario both BS densities and user densities are similar between the two MNOs, and they are both distributed as PPPs. 
This is mostly true for the rural area where the ratio between the BS densities of MNO 1 and MNO 2 (corresponding to $1.12$) is almost equal to the ratio of the respective user densities ($1.16$). The ratio of the user densities in other areas is slightly lower, resulting in a lower imbalance in the performance of the switchoff strategies between the two MNOs.

With full NS, the higher efficiency comes from the fact that the average distance between each user and its serving base station is decreased (indeed, any user has more options as to which BS to associate), thus improving the mean performance perceived by the user. Moreover, full NS allows for an overall interference reduction because users are distributed over two distinct frequency bands, effectively decreasing the average interference received by each user.

\begin{table}[t!]
\centering
\caption{Energy savings (\%) for different strategies and periods under the HLP model for the business traffic profile.}
\begin{tabular}{lccc}
\toprule
\textbf{Period} & \textbf{Full NS} & \textbf{MNO 1} & \textbf{MNO 2} \\
& & \textbf{switchoff} & \textbf{switchoff}\\
\midrule
weekday  & 33.27 & 26.44 & 27.74 \\
weekend  & 35.68 & 34.74 & 35.54 \\
\bottomrule
\end{tabular}
\label{tab:savings_business}
\end{table}

%

Finally, we have analyzed the impact of the weekday-weekend traffic patterns on the savings achievable with the different NS strategies considered.  
%
As Table \ref{tab:savings_business} shows, the weekend traffic patterns allow increasing the savings achievable with all NS schemes. Indeed, the use of NS schemes increases the degree of flexibility and adaptability of the network configuration, and the impact of this is larger the more variable and irregular the traffic pattern. Interestingly, operator switchoff strategies also perform well over weekends. This is due to the fact that lower traffic levels also lead to lower interference levels, thus decreasing the advantage that full NS has over operator switchoff sharing strategies.

\section{Conclusions}
In this work, we propose an analytical framework to evaluate the performance of QoS-aware network sharing schemes in a cellular network. 
The flexibility of the parameter space and the stochastic nature of the model make our proposed tool scalable and suitable for a wide range of scenarios.
We demonstrate its potential in providing high-level insights into the savings achievable through network sharing, by applying it to a realistic, measurement-based scenario with two MNOs.
We perform a first assessment of the impact that a set of parameters, such as peak-to-through ratio, the degree of load proportionality of the BS energy model, or the variability of traffic patterns over the week, have on the effectiveness of the most common NS strategies. Results show that the most flexible NS schemes, based on full cooperation among MNOs, 
have the potential to enable substantial energy savings with respect to QoS-aware energy optimization without sharing. 

%
%
%

\section*{Acknowledgments}
This paper was supported by the UNITY-6G project funded by the European Union’s Horizon Europe Research and Innovation Programme under Grant Agreement N. 101192650, the INTERACT COST Action, and the German Research
Foundation (DFG) within the DyMoNet project under grant
DR 639/25-1.



\end{document}